\documentclass[letterpaper,12pt,leqno]{article}
\usepackage{paper}
\usepackage[disable]{todonotes}

\newcommand{\st}{{\rm\ s.t.\ }}

\newcommand{\aalpha}{\alpha}
\newcommand{\bbeta}{\beta}

 \theoremstyle{definition}

\newcommand{\bib}{refs.bib}

\begin{document}

\newtheorem{result}{Result}
\newtheorem{example}{Example}
\newtheorem{aside}{Aside}
\newtheorem{comments}{Comment}
\newtheorem{conjecture}{Conjecture}

\newcommand{\ar}{{\rm AR}}
\newcommand{\ad}{{\rm ad}}
\newcommand{\ability}{{\rm ability}}
\newcommand{\always}{{\rm always-taker}}
\newcommand{\aand}{{\hskip1cm {\rm and}\ \ }}
\newcommand{\ave}{{\rm ave}}
\newcommand{\aven}{\frac{1}{N}\sum_{i=1}^N}
\newcommand{\alphas}{\alpha^*}
\newcommand{\atsign}{$0$}

\newcommand{\bootstrap}{\mathrm{boot}}
\newcommand{\bepsilon}{\mathbf{\varepsilon}}
\newcommand{\been}{{\bf 1}}
\newcommand{\betas}{\beta^*}
\newcommand{\betablp}{\beta_\blp}
\newcommand{\bg}{{\bf G}}
\newcommand{\bi}{{\bf I}}
\newcommand{\blp}{{\rm blp}}
\newcommand{\bp}{{\bf P}}
\newcommand{\bpx}{{{\bf P}_{\bf X}}}
\newcommand{\by}{{\bf Y}}
\newcommand{\bx}{{\bf X}}
\newcommand{\br}{{\bf R}}
\newcommand{\bw}{{\bf W}}
\newcommand{\bww}{{\bf w}}
\newcommand{\brr}{{\bf r}}
\newcommand{\ba}{{\bf A}}
\newcommand{\bz}{{\bf Z}}
\newcommand{\bv}{{\bf V}}
\newcommand{\bs}{{\bf S}}

\newcommand{\cals}{{\cal S}}
\newcommand{\calh}{{\cal H}}
\newcommand{\calt}{{\cal T}}
\newcommand{\ccc}{{\cal C}}
\newcommand{\ccb}{{\cal S}}
\newcommand{\ccr}{{\cal R}}
\newcommand{\caln}{{\cal N}}
\newcommand{\calg}{{\cal G}}
\newcommand{\calp}{{\cal P}}
\newcommand{\ccii}{{\rm CI}}
\newcommand{\ca}{\mathbb{A}}
\newcommand{\ci}{{\rm CI}}
\newcommand{\comp}{{\rm complier}}
\newcommand{\ct}{{\rm c}}
\newcommand{\cluster}{{\rm cluster}}
\newcommand{\combined}{{\rm comb}}

\newcommand{\data}{\mathrm{data}}
\newcommand{\define}{:=}
\newcommand{\did}{\mathrm{did}}
\newcommand{\donald}{\mathrm{Donald}}
\newcommand{\dif}{{\rm dif}}

\newcommand{\emphunderline}{\underline}
\newcommand{\earn}{{\rm earnings}}
\newcommand{\educ}{{\rm educ}}
\newcommand{\exper}{{\rm exper}}
\newcommand{\expers}{{\rm exper}^2}
\newcommand{\el}{{\rm el}}
\newcommand{\ehw}{{\rm EHW}}
\newcommand{\ehww}{{\rm ehw}}

\newcommand{\ff}{{\rm f}}
\newcommand{\fs}{{\rm fs}}
\newcommand{\frd}{{\rm frd}}
\newcommand{\finsin}{\frac{1}{N}\sum_{i=1}^N}
\newcommand{\fin}{\frac{1}{N}}

\newcommand{\gary}{\mathrm{Gary}}
\newcommand{\gls}{{\rm fgls}}
\newcommand{\gs}{g}
\newcommand{\GG}{G}

\newcommand{\hmmv}{{\hat{\mathbb{V}}}}
\newcommand{\high}{{\rm high}}
\newcommand{\hct}{{\rm HC2}}
\newcommand{\hcth}{{\rm HC3}}
\newcommand{\homo}{\mathrm{homo}}
\newcommand{\hatbetaols}{\hat\beta^{\rm ols}}
\newcommand{\hatthetagmm}{\hat\theta_{\rm gmm}}
\newcommand{\hatthetael}{\hat\theta_{\rm el}}
\newcommand{\hatthetaml}{\hat\theta_{\rm ml}}
\newcommand{\htau}{\hat{\tau}}

\newcommand{\iv}{\mathrm{IV}}
\newcommand{\indep}{\perp\!\!\!\perp}

\newcommand{\josh}{\mathrm{Josh}}
\newcommand{\jack}{\mathrm{jacknife}}

\newcommand{\lambdas}{\lambda^*}
\newcommand{\lb}{{\rm lb}}
\newcommand{\liml}{{\rm liml}}
\newcommand{\learn}{{\rm log(earnings)}}
\newcommand{\lectures}{{Lectures in Econometrics,\ }}

\newcommand{\mrelec}{\mathrm{elec}}
\newcommand{\mrcap}{\mathrm{cap}}
\newcommand{\mroper}{\mathrm{oper}}
\newcommand{\mrno}{\mathrm{no}}
\newcommand{\mrgas}{\mathrm{gas}}
\newcommand{\medd}{{\rm med}}
\newcommand{\mm}{{\rm m}}
\newcommand{\med}{{\rm med}}
\newcommand{\mle}{{\rm mle}}
\newcommand{\mmr}{{\mathbb{R}}}
\newcommand{\mmw}{{\mathbb{W}}}
\newcommand{\mmz}{\mathbb{Z}}
\newcommand{\mx}{\mathbb{X}}
\newcommand{\ml}{{\rm ml}}
\newcommand{\mme}{{\mathbb{E}}}
\newcommand{\mmv}{{\mathbb{V}}}
\newcommand{\mma}{{\mathbb{A}}}
\newcommand{\mmva}{{\mathbb{AV}}}
\newcommand{\mmc}{{\mathbb{C}}}
\newcommand{\mmx}{{\mathbb{X}}}
\newcommand{\modrobust}{{\rm HC2}}
\newcommand{\mvar}{\mathbb{V}}
\newcommand{\mmav}{{\mathbb{AV}}}

\newcommand{\nfn}{N_{{\rm f}\ct}}
\newcommand{\nmn}{N_{{\rm m}\ct}}
\newcommand{\nfe}{N_{{\rm f}\tc}}
\newcommand{\nme}{N_{{\rm m}\tc}}
\newcommand{\never}{{\rm never-taker}}
\newcommand{\nc}{N_{\ct}}
\newcommand{\nt}{N_{\tc}}
\newcommand{\nf}{N_{\ff}}
\newcommand{\nm}{N_{\mm}}
\newcommand{\neyman}{\mathrm{neyman}}

\newcommand{\opsn}{o_p\left(N^{-1/2}\right)}
\newcommand{\Opsn}{O_p\left(N^{-1/2}\right)}
\newcommand{\opn}{o_p\left(N^{-1}\right)}
\newcommand{\Opn}{O_p\left(N^{-1}\right)}
\newcommand{\obs}{{\rm obs}}
\newcommand{\oy}{\overline{Y}}
\newcommand{\orr}{\overline{R}}
\newcommand{\ols}{{\rm ols}}

\newcommand{\pop}{{\rm pop}}
\newcommand{\pate}{{\rm pate}}
\newcommand{\pos}{{\rm pos}}
\newcommand{\popt}{{\rm patt}}
\newcommand{\pr}{{\rm pr}}

\newcommand{\rank}{{\rm rank}}
\newcommand{\reg}{{\rm reg}}
\newcommand{\robust}{{\rm robust}}
\newcommand{\ress}{Y_i-X_i hatbetaols}
\newcommand{\rmelec}{{\rm elec}}
\newcommand{\relec}{{\rm elec}}
\newcommand{\rmoper}{{\rm oper}}
\newcommand{\rmcap}{\mathrm{cap}}
\newcommand{\rmgas}{\mathrm{gas}}
\newcommand{\rmno}{\mathrm{no}}
\newcommand{\rmif}{{\rm if}}

\newcommand{\qob}{{\rm qob}}

\newcommand{\subs}{\mathrm{subsampling}}
\newcommand{\sate}{{\rm sate}}
\newcommand{\sample}{{\rm sample}}
\newcommand{\samplet}{{\rm satt}}
\newcommand{\spp}{{\rm sp}}
\newcommand{\srd}{{\rm srd}}
\newcommand{\se}{{\rm s.e.}}
\newcommand{\strata}{{\rm strat}}
\newcommand{\snn}{\sum_{i=1}^N}
\newcommand{\snt}{\sum_{t=1}^T}
\newcommand{\str}{^{*}}

\newcommand{\tc}{{\rm t}}
\newcommand{\tsls}{{\rm tsls}}
\newcommand{\tmle}{\hat\theta^{\rm mle}}
\newcommand{\thetaml}{\theta^{\rm mle}}
\newcommand{\ty}{\tilde Y}
\newcommand{\tick}{\checkmark}
\newcommand{\thetas}{\theta^*}
\newcommand{\taup}{\tau_\spp}
\newcommand{\tautp}{\tau_{\spp,\tc}}

\newcommand{\ow}{\overline{W}}
\newcommand{\oz}{\overline{Z}}
\newcommand{\ox}{{\overline{X}}}

\newcommand{\ub}{{\rm ub}}

\newcommand{\veen}{{\rm homo}}
\newcommand{\vtweea}{{\rm    homo,unbiased}} 
\newcommand{\vtwee}{{\rm homo,unbiased,t-dist}}
\newcommand{\vdrie}{{\rm robust}} 
\newcommand{\vet}{{\rm veteran}}

\newcommand{\wb}{{\bf W}}
\newcommand{\win}{W_{i}}
\newcommand{\welch}{\mathrm{welch}}

\newcommand{\xbj}{{\bf X}_{(j)}}
\newcommand{\xb}{{\bf X}}
\newcommand{\xbi}{{\bf X}_{(i)}}
\newcommand{\xxb}{{\bf x}}

\newcommand{\yoin}{Y_{i}}
\newcommand{\yin}{Y_i(0)}
\newcommand{\yie}{Y_i(1)}
\newcommand{\yinn}{Y_{i}(0)}
\newcommand{\yien}{Y_{i}(1)}
\newcommand{\ybn}{{\bf Y}(0)}
\newcommand{\ybe}{{\bf Y}(1)}
\newcommand{\yb}{{\bf Y}}
\newcommand{\ybo}{{\bf Y}^{\rm obs}}
\newcommand{\ybm}{{\bf Y}^{\rm mis}}
\newcommand{\yybn}{{\bf y}(0)}
\newcommand{\yybe}{{\bf y}(1)}
\newcommand{\ybni}{{\bf Y}_{(i)}(0)}
\newcommand{\ybei}{{\bf Y}_{(i)}(1)}
\newcommand{\ybnj}{{\bf Y}_{(j)}(0)}
\newcommand{\ybej}{{\bf Y}_{(j)}(1)}
\newcommand{\yoi}{Y^{\rm obs}_i}

\newcommand{\zin}{Z_{i}}

\title{Identification of Average Treatment Effects in Nonparametric Panel Models}
\author{Susan Athey \& Guido Imbens
\thanks{Susan Athey, Stanford University; Guido Imbens, Stanford University.
We thank Peter Bickel, Kevin Chen, Harold Chiang, Peng Ding, Avi Feller,  Apoorva Lal, Lihua Lei, and Yiqing Xu
 for helpful comments and discussions. 
We thank
the Office of Naval Research for support under grant numbers N00014-17-1-2131 and N00014-19-1-2468 and Amazon for a gift.
}}
\date{March 2025}                       
\begin{titlepage} \maketitle This paper studies identification of average treatment effects in a panel data setting. It introduces a novel nonparametric factor model and proves identification of average treatment effects. The identification proof is based on the introduction of a consistent estimator.  Underlying the proof is a result that there is a consistent estimator for the expected outcome in the absence of the treatment for each unit and time period; this result can be applied more broadly, for example in problems of decompositions of group-level differences in outcomes, such as the much-studied gender wage gap.
\end{titlepage}

\listoftodos

\section{Introduction}\label{s:introduction}

There is a large literature on estimating causal effects and predictive estimands using panel data. Various estimators have been proposed, often motivated by different models of the underlying data generating process. There is not a widely used model with an accompanying set of assumptions that is the focal point of this literature, a situation that contrasts with the cross-section setting, where there are several commonly used sets of assumptions such as unconfounded treatment assignment and overlap in the distribution of propensity scores. The panel data literature most commonly relies on estimators that are justified at least in part by functional form assumptions, but the precise role of these assumptions has not been fully described. In this paper we propose a model of the data generating process together with a set of conditions under which the conditional expectations of outcomes given unobserved unit and time components are identified, where the conditions do not include functional form restrictions. Thus, our approach is analogous in spirit to the approach taken in the literature on identification of causal effects under the unconfoundedness and overlap assumptions (see e.g. \cite{imbens2015causal}). We the extend these results for predictive estimands to the identification of average causal effects and decomposition estimands.

The outcome model  is a generative structural model that takes the form of a Nonparametric Panel Model (NPM):
\begin{equation}\label{eq:modelintro} Y_{it}(0)=g(\alpha_i,\beta_t,\varepsilon_{it}),\hskip5cm {\bf (NFM)},\end{equation}
where the unobserved components  $\aalpha_i$, $\beta_t$ and $\varepsilon_{it}$ may be vector-valued.
Our main results rely on several assumptions about these components. First, the sets $\{\aalpha_i\}_{i=1}^{N},$  
$\{\bbeta_t\}_{t=1}^{T},$ and $\{\varepsilon_{it}\}_{i,t},$ are independent of each other. In addition, the $\aalpha_i$ are independent across units. However, the $\beta_t$ and $\varepsilon_{it}$ may be autocorrelated over time,  but we assume stationarity of the process. 
These assumptions are satisfied if $g$ has an additive form, as in the popular Two-Way-Fixed-Effect (TWFE) models, or if it takes the form of a linear factor model. However, much richer models are consistent with this setup.

Although we cannot identify the unobserved unit and time components $\aalpha_i$ and $\bbeta_t$ under our assumptions, we can both identify and consistently estimate the expected value $\mu(\aalpha_i,\bbeta_t)=\mathbb{E}[Y_{it}|\aalpha_i,\bbeta_t]$, which is a key building block for both our average treatment effect  and decomposition results. 
Beyond the generative model, this result relies solely on smoothness and restrictions on the asymptotic sequences.
We do so by constructing sets of units ${\cal J}(i)\subset\{1,\ldots,N\}$ with increasing cardinality such that for units $j\in{\cal J}(i)$ in this set the function $\mu(\alpha_j,\bbeta)$ is similar to $\mu(\aalpha_i,\beta)$ as a function of $\bbeta$. We construct these sets by finding units $j$ such that, for all other units $k$, the covariance over time between $Y_{it}$ and $Y_{kt}$ is similar to the covariance between $Y_{jt}$ and $Y_{kt}$. We then estimate $\mu_{it}$ by averaging $Y_{jt}$ for units in such sets ${\cal J}(i).$

For the causal problem,
we focus on a setting with a binary treatment $W_{it}$ that affects only unit $i$ in period $t$ (thus ruling effects of the treatment that persist over time). This allows the potential outcomes to be defined in terms of the contemporaneous treatment, $Y_{it}(0)$ and $Y_{it}(1)$.
We focus on identification of the average treatment effect for the treated,
\begin{equation}\label{eq:ATT} \tau\equiv \frac{1}{\sum_{i,t} W_{it}}\sum_{i,t} W_{it}\Bigl(Y_{it}(1)-Y_{it}(0)\Bigr),\hskip3cm {\bf (ATT)},\end{equation}
although our results can be generalized to other average treatment effects.
The primary condition for our result is that conditional on $\aalpha_i$ and $\bbeta_t$, assignment to treatment is independent of $Y_{it}(0)$:
\begin{equation}\label{eq:assignment}  W_{it}\indep\ Y_{it}(0)\  \Bigl|\ \aalpha_i,\bbeta_t ,\hskip2cm {\bf (Latent\ Factor\ Unconfoundedness)}.\end{equation} In addition we need an overlap condition that the probability of receiving the treatment conditional on $\alpha_i$ and $\beta_t$ is bounded away from zero. This Latent Factor Unconfoundedness assumption is analogous to unconfoundedness in the cross-section setting.
A final condition is that for $\mu(\alpha,\beta)\equiv \mathbb{E}[Y_{it}|\alpha_i=\alpha,\beta]$, $\mathbb{E}_\beta[\mu(\alpha,\beta)-\mu(\alpha',\beta))^2]=0$ implies $\alpha=\alpha'$.
Our main result for the treatment effect setting is that given sufficient smoothness on $g(\cdot)$,
the average effect for the treated $\tau$ is identified in this setting when both $T$ and $N$ go to infinity.

In addition to causal problems, the methods may also be applied to other purely predictive problems that rely on identification and estimation of $\mu_{it}$. One such category of problems is decomposition of a difference in outcomes (e.g. by demographic group or other categorization) into explained and unexplained components, as is common in the literature on gender wage gaps \cite{blau2017gender}. We develop this application further in Section \ref{sec:decomp}. 

\section{Related Literature}

There are a number of distinct literatures that the current paper builds on.
One is the econometric literature on panel data. Much of this literature has focused on Two-Way-Fixed-Effect (TWFE) models
and related Difference-In-Differences (DID) estimators, and more recently linear factor models \cite{bai2009panel}. See 
 \citep{chamberlain1984panel, arellano2001panel, arellano2003panel,baltagi2008econometric,hsiao2022analysis, wooldridge2010econometric,arellano2011nonlinear} for textbook discussions 
 and \cite{arkhangelsky2024causal,roth2022s} for recent surveys.

Extensions allowing for nonlinear functions of linear factor models are considered in \cite{chen2021nonlinear, yalcin2001nonlinear, feng2020causal, feng2023optimal, freyberger2018non, zeleneev2020identification, abadiecausal}.

The current study is also related to the literature on synthetic control methods, \cite{abadie2003,abadie2010synthetic, abadie2019using, arkhangelsky2021synthetic, liu2020practical, fry2024method}. In that literature  the generative models are often not specified explicitly, with the focus on algorithms that lead to effective estimators.

Another literature that is relevant is the research on row and column exchangeable arrays, see \cite{aldous1981representations, lynch1984canonical, 
mccullagh2000resampling}. Nonparametric models motivated by the double exchangeability have been considered in network settings, where they are referred to as graphons.
The focus in this literature is on estimating the conditional probability of links between nodes in the network, assuming the probability depends in an unrestricted way on unit specific components for both nodes.
See
for key insights
\cite{bickel2009nonparametric, bickel2011method, zhang2017estimating, graham2024sparse, chiang2023inference}.

Very relevant is also the work by
\cite{bonhomme2015grouped, bonhomme2022discretizing, cytrynbaum2020blocked} on grouped heterogeneity. In this line of work units are grouped through k-means clustering so that units in the same group have the same values for the unobserved unit components.



\section{Set Up}\label{section:setup}

We are interested in estimating causal effects in a panel data setting. We focus on a setting with a binary treatment, with $\bw$ the $N\times T$ matrix of treaments, with typical element $W_{it}\in\{0,1\}$. There are two matrices of potential outcomes
$\by(0)$ and $\by(1)$.  We observe the matrix of realized outcomes $\by$ with typical element $Y_{it}=Y_{it}(W_{it})$. This notation already imposes the stable unit treatment values assumption or SUTVA \cite{rubin1978bayesian, imbens2015causal},  ruling out dynamic effects.

One of the estimands that is a main focal point of the current study is the average effect on the treated,
\begin{equation}
    \tau\equiv 
    \sum_{i,t} W_{it} \Bigl(Y_{it}(1)-Y_{it}(0)\Bigr)
    \Bigl/
    \sum_{i,t} W_{it} ,\hskip5cm {\bf (ATT)}
\end{equation}
\begin{remark}
The focus on the average effect for the treated is for ease of exposition and not essential for the ideas developed in this paper. Similar results hold for the overall average effect.\end{remark}
\begin{remark}
In most of the discussion we abstract from the presence of exogenous covariates.
\end{remark}

Our starting point, and a key contribution, is a fully nonparametric generative model that captures the relation between  outcomes for different units and time periods:
\begin{assumption}\label{assumption:model}
The control outcomes satisfy
\begin{equation}\label{eq:model} Y_{it}(0)=g(\alpha_i,\beta_t,\varepsilon_{it}),\hskip5cm {\bf (NFM)}\end{equation}
with\\
\\ $(i)$ 
 $\alpha_i$, $\beta_{t}$ and $\varepsilon_{it}$ are finite dimensional, 
 \\ $(ii)$   the $\{\alpha_i\}_{i=1}^N$ are $\{\beta_{t}\}_{t=1}^T$ and $\{\varepsilon_{it}\}_{i,t}$ are jointly independent. In addition the $\alpha_i$ are independent across units, although the $\beta_t$ and $\varepsilon_{it}$ can be autocorrelated, although they need to have stationary distributions.
\end{assumption}
\begin{remark}
One special case and leading example of this structure in empirical work is the Two-Way-Fixed-Effect (TWFE) model, where the function $g(\cdot)$ satisfies
\begin{equation}\label{eq:twfe} g(\aalpha_i,\bbeta_t,\varepsilon_{it})=\aalpha_i+\bbeta_t+\varepsilon_{it},\end{equation}
often with the assumption that $\varepsilon_{it}$ is mean zero, homoskedastic, and independent over time.

A second special case is the  Linear Factor Model (LFM), where
\begin{equation}\label{eq:linearfactor} g(\aalpha_i,\bbeta_t,\varepsilon_{it})=\aalpha_i^\top\bbeta_t+\varepsilon_{it}.\end{equation}
\end{remark}
\begin{remark}
In both the TWFE and LFM literature the $\alpha_i$ and $\beta_t$ are typically viewed as parameters to be estimated, with assumptions made only about the stochastic properties of the $\varepsilon_{it}$ conditional on any sequence of values for $\aalpha_i$ and $\bbeta_t$. Here we treat the $\alpha_i$ and $\beta_t$ explictly as random variables and make assumptions about their properties. These assumptions do not restrict their distributions, nor do they impose restrictions on the association with covariates in settings when we include those later,  in the spirit of Chamberlain's correlated random effects  model \cite{chamberlain1984panel}.
\end{remark}

\begin{remark}
For any given size matrix $\by$, with $N$ rows and $T$ columns, one can write that matrix as a rank $\min(T,M)$ linear factor model
$Y_{it}=\alpha_i^\top\beta_t$ without any $\varepsilon_{it}$. One may be able to approximate $\by$ using a LFM with substantially fewer factors. This suggest that a linear factor model in itself need not be  a restrictive structure. Why then, do we wish to consider nonparametric/nonlinear factor models? One reason is that although the approximation works for a given matrix, the same number of  factors and factor loadings need not  be sufficient for larger $N$ and $T$, making the representation less attractive as a generative model. To formalize this discussion we later introduce a sequence of populations with restrictions on the sequence of models.    
\end{remark}

\begin{remark}
The generative TFM model  can be motivated by
 row and column exchangeability of $\by,$ meaning that the distribution of $\by$ is not changed by arbitrarily permuting the time or unit indices.
 See for a general discussion on exchangeability \citep{de2017theory}, and for definitions in settings with arrays  \citep{aldous1981representations, lynch1984canonical, 
mccullagh2000resampling}.
If $\by(0)$ is row and column exchangeable, then
there is a representation
\[ Y^*_{it}=g(\alpha_i,\beta_t,\varepsilon_{it}),\]
with $g(\cdot)$ measurable, 
such that the distribution of 
$\by^*$ is the same as that of $\by$, with $\alpha_i$, $\beta_t$, and $\varepsilon_{it})$ all scalar,  jointly independent, and uniformly distributed.
\end{remark}
\begin{remark}
Note that the exchangeability result implies there is a function $g(\cdot)$ in the representation that has scalar arguments. We allow in the TFM model in (\ref{eq:model})
to have vector-valued arguments. At this stage allowing for vector-valued components does not add generality. However, once we make smoothness assumptions on the function $g(\cdot)$ beyond measurability, this may affect the quality of the approximations. In addition, allowing for vector-valued components facilitates comparisons to the linear factor model in (\ref{eq:linearfactor}).
\end{remark}

To be able to be precise about consistency we first need to be precise about the sequences of matrices  or populations. Let us index the sequence of matrices by $m=1,2,\ldots.$ The dimensions of the matrix $\by_m$ are $N_m$ and $T_m$.
\begin{assumption}\label{assumption:asymptotic}{\sc (Asymptotic Sequences)}\\
$(i)$    $N_m,T_m\rightarrow \infty$,\\
$(ii)$ 
$g_m(\cdot)$ is identical for all $m$ (so we can omit the subscript $m$).
\end{assumption}

\begin{remark} There is some important content to this assumption although there is nothing testable. We cannot  deal in general with the case where we have $g(\cdot)$ indexed by the population, for example, $g_m(\aalpha,\bbeta,\varepsilon)=h(m\aalpha,\bbeta,\varepsilon)$. In that case, for any given value of $\aalpha_1$, we would not be guaranteed to have a lot of units $i$ that have a value $m\aalpha_j$ that is close to $m\aalpha_1$. So this assumption about the asymptotic sequences captures the idea that as we get more units and more time periods they are filling in the space, not purely expanding it, similar to the way infill asymptotics is utilized in time series analyses and spatial econometrics \cite{zhang2005towards}.
\end{remark}
We also need some smoothness conditions. Note that the row and column exchangeability implies the existence of a measurable function $g_m(\cdot)$.
In the smoothness assumption we strengthen  that to include continuity and Lipschitz conditions.
\begin{assumption}\label{assumption:smoothness}{\sc (Smoothness Condition)}\\
$g(\aalpha,\bbeta,\varepsilon)$ is continuously differentiable in all its arguments with first derivatives bounded.
\end{assumption}

\begin{assumption}\label{assumption:rank}{\sc (Rank Condition)}\\
If 
$\mathbb{E}_\beta[(\mu(\alpha,\beta)-\mu(\alpha',\beta))^2]=0$, then $\alpha=\alpha'.$
\end{assumption}

\section{Identification of the Conditional Means}

Before focusing on the average treatment effect or decomposition results, we show a preliminary identification result for the expected value of $Y_{it}(0)$ given $\alpha_i$ and $\bbeta_t$. This is a key technical result that is a building block for the more substantively interesting identification results discussed later.
For this result we focus on a predictive setting where we observe the entire matrix $\by(0)$, and do not use the Latent Factor Unconfoundedness assumption.
Clearly identification of the entire function $g(\cdot)$, or of the unit and time components $\aalpha_i$ and $\bbeta_t$ is not feasible based on Assumptions \ref{assumption:model}, \ref{assumption:asymptotic}, and \ref{assumption:smoothness} without more structure. 
However, we do not need to estimate either the function $g(\cdot)$ itself or  its arguments. For estimating average treatment effects and for our decomposition results it suffices to estimate the conditional expectation of $g(\aalpha_i,\bbeta_t,\varepsilon_{it})$, conditional on $(\aalpha_i,\bbeta_t)$, for all the treated  pairs of indices $(i,t)$. 
Define the conditional expectations
\[ \mu(\aalpha,\bbeta)\equiv \mathbb{E}[g(\aalpha,\bbeta,\varepsilon)|\aalpha,\bbeta],\qquad \mu_{it}\equiv \mu(\alpha_i,\beta_t)
\]
and the residual functions
\[ \eta(\aalpha,\bbeta,\varepsilon)\equiv g(\aalpha,\bbeta,\varepsilon)-\mu(\aalpha,\bbeta).\qquad\eta_{it}\equiv\eta(\alpha_i,\bbeta_t,\varepsilon_{it}).\]

In this section we focus on consistently estimating $\mu_{it}$ for a single  unit/period pair $(i,t)$ as a building block. So we want to have an estimator $\hat Y_{it}$ that is a function of $\by$ such that
\begin{equation}\label{eq:consistency} \hat{Y}_{it}-\mu_{it}\stackrel{p}{\longrightarrow} 0,\end{equation}
under Assumptions \ref{assumption:model}, \ref{assumption:asymptotic}, and \ref{assumption:smoothness}.

We  establish this identification result by  constructing a set ${\cal J}_m(i)\subset \{1,2,\ldots,N_m\}$ with two properties under  Assumptions \ref{assumption:model}, \ref{assumption:asymptotic}, and \ref{assumption:smoothness} (we do not need Assumption \ref{assumption:unconfoundedness} for this discussion): $(i)$  units $j$ in this set have $\mu(\alpha_j,\bbeta)$ close to $\mu(\alpha_i,\bbeta)$ for all $\bbeta$ with high probability, and $(ii)$ the cardinality of the set increases without bounds with $m$.
(And of course a similar strategy would be to look for the corresponding set of time indices.)
The key insight is that in order to check that unit $j$ and unit $i$ have similar values of the function $\mu(\alpha_i,\bbeta)$ and $\mu(\aalpha_j,\bbeta)$ for all $\bbeta$, it is {\it not} sufficient to compare outcomes $Y_{it}$ and $Y_{jt}$ for units $i$ and $j$ over time. Instead we compare their covariances with other units. Specifically, we compare the covariance between unit $i$ and unit $k$ with the covariance between unit $j$ and unit $k$. If for large $T_m$ and $N_m$ those covariances are similar for {\it all} other comparison units $k$, then it must be the case that  $\mu(\alpha_i,\bbeta)$ and $\mu(\alpha_j,\beta)$ are similar for all $\bbeta$. Note that it does {\it not} mean that $\alpha_i$ and $\alpha_j$ are close, but it means that we can use unit $j$ for estimating the conditional mean for unit $i$ in the same period.
A second insight is that although we do not directly observe $\mu(\aalpha_i,\bbeta)$ for unit $i$, we can
estimate the covariance between $\mu(\aalpha_i,\bbeta)$ and 
$\mu(\aalpha_j,\bbeta)$ using the covariance between $Y_{it}$ and $Y_{kt}$ over time.

\begin{remark} Three natural approaches for constructing  sets ${\cal J}_m(i)$ with the two aforementioned properties do not work.

The first is a simple matching strategy where for a given unit $i$ we look for the closest units $j$ in terms of the average value over time, $\overline{Y}_{i\cdot}=\sum_{t=1}^{T_m} Y_{it}/T_m$. This strategy works in a TWFE setting but not in more general settings. 
To see why this does not work in general suppose $\aalpha_i$ and $\bbeta_j$ are both uniformly distributed on $[0,1]$ and $g(\kappa,\aalpha,\bbeta,\varepsilon)=\aalpha(\bbeta-1/2)$ so that $\mu(\aalpha)=0$ for all $\aalpha$.

The second approach is also a matching approach. Suppose we look for units $j$ that are the closest to unit $i$, that is, units that  minimize $\sum_{t=1}^{T_m} (Y_{jt}-Y_{it})^2/T_m$. Although this approach may work in additive settings where $\eta(\aalpha_i,\bbeta_t,\varepsilon_{it})$ does not depend on $\aalpha_i,\bbeta_t)$, it does not work in general. To see that this does not work in general, consider an example where $g(\kappa,\aalpha,\bbeta,\varepsilon)=\aalpha(\beta+\varepsilon)$, with $\aalpha$, $\beta$ and $\varepsilon$  ${\cal N}(0,1)$ and $\varepsilon$ has a ${\cal N}(0,1)$ distribution. Suppose the treated unit $i$ has $\aalpha_i=1$. Then for a unit $j$ with $\aalpha_j=\aalpha_i=1$, the expected value of $\sum_{t=1}^{T_m} (Y_{jt}-Y_{it})^2/T_m$ is equal to $\mathbb{E}[(\varepsilon_{it}-\varepsilon_{jt})^2]=2.$  In that case unit $i$ will get matched with units $j$ with $\aalpha_j\approx 1/2$.

A third approach starts with the observation that if $\aalpha_j$ is approximately equal to $\aalpha_i$, it must be the case that the distribution of $Y_{it}$ over time identical to the distribution of $Y_{jt}$ over time. Let $F_i(y)$ denote the cumulative distribution function of $Y_{it}=g(\aalpha_i,\beta_t,\varepsilon_{it})$ conditional on  $\aalpha_i$. A strategy based on this would look for $j$ such that
$\sup_y |F_i(y)-F_j(y)|$ is close to zero. (In practice of course one would need to use estimated versions of these cumulative distribution functions, but with $T_m$ large one could estimate them precisely.) To see that this strategy does not work, consider the case where $g(\aalpha,\bbeta,\varepsilon)=(\aalpha-1/2)^2(\bbeta-1/2)$, with both $\aalpha_i$ and $\bbeta_t$ uniformly distributed on $[0,1]$. Consider a unit $i$ with $\aalpha_i=0$. The cumulative distribution function for such a unit is the same as for a unit $j$ with $\aalpha_j=1$.

Note that if the TWFE model holds, then all three of the aforementioned strategies do work. However, outside of the TWFE model these strategies do not generally work.
These three examples of failed strategies show that a challenge is finding a metric that differentiates between different values of the unit (or time) component that lead to different functions $\mu(\aalpha_i,\bbeta_t)$. Our strategy will focus on metrics that involve multiple units rather than simply focus on the marginal distribution of the outcome for a specific unit. 
\end{remark}

First we define an infeasible version of the set  ${\cal J}_m(i)$ that helps develop the intuition for the identification result.
Define
\[ {\cal J}^*(\aalpha)\equiv 
\left\{
\aalpha'\in[0,1],\st 
\sup_{\alpha''}
\biggl|\mathbb{E}_\beta\left[\left.\left\{\mu(\aalpha,\bbeta)-\mu(\aalpha',\bbeta)\right\}\mu(\aalpha'',\bbeta)\right|\aalpha,\aalpha',\aalpha''\right]
\biggr|=0
\right\}
.\]
\begin{lemma}\label{lemma_main}
If $\aalpha'\in{\cal J}^*(\aalpha)$ then $\mathbb{E}_\beta[\{\mu(\alpha,\beta)-\mu(\alpha',\beta)\}^2|\aalpha,\aalpha']=0.$   
\end{lemma}
\begin{remark}
If $\alpha'\in{\cal J}^*(\aalpha)$ then $\mu(\aalpha,\bbeta)$ is close to $\mu(\aalpha',\bbeta)$ for all $\bbeta.$ It is not, however, the case that $\aalpha$ and $\aalpha'$ are necessarily close.    
\end{remark}
\begin{remark}
The  expression $\sup_{\alpha''}
|\mathbb{E}_\beta[\{\mu(\aalpha,\bbeta)-\mu(\aalpha',\bbeta)\}\mu(\aalpha'',\bbeta)|\aalpha,\aalpha',\aalpha'']
|$ is a $L_\infty$ version of the similarity distance \cite{lovasz2010regularity, lovasz2012large}.
\end{remark}
\begin{remark}
We only need the covariances to be the same for $\aalpha''\in\{\aalpha,\aalpha'\}$. However, because we do not observe the values, we end up having to check this for all values of $\aalpha''\in[0,1].$    
\end{remark}

\begin{remark}
It is interesting to consider the set ${\cal J}^*(\aalpha)$ in the TWFE and linear factor model settings.
\end{remark}

We cannot use this first infeasible identification result directly to create, for a given unit $i$,  a set of units with similar functions $\mu(\aalpha,\bbeta)$ because we do not know either the values of the $\aalpha_i$ nor the function $\mu(\aalpha,\bbeta)$. This leads to three  complications and corresponding modifications of the set ${\cal J}^*(\aalpha)$ to turn this into an actual identification result. First, we compare the covariances only at the sample values $\aalpha_i$, still assuming we know the entire functions $\mu(\aalpha_i,\bbeta)$ as a function of $\bbeta$, but now only for all the realized values $\aalpha_i$. This implies that we cannot find pairs of units with exactly the same $\mu(\aalpha_j,\bbeta)$, only approximately so, with the apporoximation improving as the number of units $N_m$ increases. Second, we only evaluate the $\mu(\aalpha,\bbeta)$ at the sample values of $\bbeta_t$, instead of taking the expectation over $\beta_t$. In other words, we focus on the average $\sum_t \mu(\aalpha_i,\bbeta_t)\mu(\aalpha_k,\bbeta_t)/T_m$ being close to the average  $\sum_t \mu(\aalpha_i,\bbeta_t)\mu(\aalpha_k,\bbeta_t)/T_m$, for all $k$.
The quality of the approximation this induces relies on $T_m$ being large. Third, we do not observe the  values of the products $\mu(\aalpha_i,\bbeta_t)\mu(\aalpha_k,\bbeta_t)$ for pairs of units $i$ and $k$ at all $t$, we need to estimate these products using the averages $\sum_{t=1}^{T_m} Y_{it}Y_{kt}/T_m$. For this we rely on $T_m$ being large relative to the number of units $i$ and $k$ where we evaluate these averages to ensure those averages approximate the expectations accurately.


The first of our  main identification results presents a feasible version of Lemma \ref{lemma_main}. It constructs a set of indices ${\cal J}_{m,\nu}(i)\subset\{1,\ldots,N_m\}$. The set  is indexed by $\nu$ which governs how close the functions $\mu(\aalpha_j,\beta)$ are to $\mu(\aalpha_j,\bbeta)$ for $j\in{\cal J}_{m,\nu}(i) $, with high probability.
Define
\[ {\cal J}_{m,\nu}(i)\equiv \left\{j=1,\ldots,N_m,\st
\max_{k\neq i,j}
\left|\frac{1}{T_m}\sum_{t=1}^{T_m} (Y_{it}-Y_{jt}) Y_{kt}\right|\leq \nu
\right\}.\]

Let $\overline{\mu}$ be an upper bound on the absolute value of $\mu(\alpha,\beta)$, and let $\overline{\mu'}$ be an upper bound on the absolute value of the derivatives of $\mu(\aalpha,\bbeta)$ with respect to $\aalpha$ and $\bbeta.$
Let $C_Y$ be an upper bound on the absolute value of $g(\alpha,\beta,\varepsilon).$

\begin{theorem}
\label{theorem1}
Suppose Assumptions \ref{assumption:model}, \ref{assumption:asymptotic}, and \ref{assumption:smoothness}

hold. For any $\xi,\epsilon>0$, if 
\[N_m\geq\frac{\ln(\epsilon\xi/(16\overline{\mu'}\overline{\mu})}{\ln(1-\xi/16\overline{\mu'}\overline{\mu})},\qquad{\rm and}\quad T_m\geq \frac{256N_m^2 C^2_Y}{\epsilon^2\xi^2},\]
then $(i)$
\[ \pr\left(\mathbb{E}_\beta\left[\left.\left\{\mu(\alpha_i,\beta)-\mu(\alpha_j,\beta)\right\}^2\right|\aalpha_i,\aalpha_j,j\in {\cal J}_{m,\xi/8}(i) \right]>\xi\right)<\varepsilon,\]
and $(ii)$ the cardinality of the set ${\cal J}_{m,\xi/8}(i)$ diverges for all $i$.
\end{theorem}


\section{Estimating Average Treatment Effects}
\label{sec:ATE}

In this section we use the results from the previous section to estimate average treatment effects under an unconfoundedness type assumption. We start from a slightly different point though, without the generating model $g(\aalpha_i,\bbeta_t,\varepsilon_{it})$.

First, we postulate the existence of a pair of potential outcomes $Y_{it}(0)$ and $Y_{it}(1),$ corresponding to the outcomes without and given the intervention. There is a binary treatment $W_{it}\in\{0,1\}$, with the realized outcome corresponding to the potential outcome given the treatment received,
\[ Y_{it}\equiv 
\left\{
\begin{array}{ll}Y_{it}(0)
\hskip1cm & {\rm if}\ W_{it}=0,\\
Y_{it}(1)& {\rm otherwise.}
\end{array}
\right.\]
We are interested in the average treatment effect for the treated,
\[ \tau\equiv \frac{1}{\sum_{i,t} W_{it}}\sum_{i,t} W_{it} \Bigl(Y_{it}(1)-Y_{it}(0)\Bigr)\qquad {\bf (ATT)}\]
The insights extend to other estimands such as the overall average treatment effect.

In order to identify $\tau$ we need some restrictions on the assignment mechanism. Our key assumption  is closely related to the standard unconfoundedness/ignorability assumptions in the cross-section evaluation literature \cite{rosenbaum1983central,  imbens2015causal}. Such assumptions are typically stated as independence of the assignment and the set of potential outcomes conditional on observed confounders. Here the critical assumption is formulated differently in two aspects. The first, minor, difference  is that we only  use {\it weak} unconfoundedness, where we assume independence of the assignment and the control potential outcome, as introduced in
\cite{imbens2000}. Second, the conditioning is on {\it unobserved} confounders, rather than observed confounders. In particular, we assume there are latent unobserved confounders such that conditional on those unconfounders assignment would be as good as random. That by itself is without loss of generality, in other words such an assumption does not have any direct content. Our assumption strengthens this by assuming that this set of latent confounders can be separated into some that are unit-specific and some that are time-specific, and {\it none} that are unit-time specific.

\begin{assumption}\label{assumption:unconfoundedness}{\sc (Latent Factor Ignorability)}\\
There are unobserved unit components $\aalpha_i$ and unobserved time components $\bbeta_t$  such that
the assignment mechanism satisfies  $(i)$ (Latent Factor Unconfoundedness):
\[ W_{it}\ \indep\ Y_{it}(0)\ \Bigl|\ \aalpha_i,\bbeta_t,\]
and 
$(ii)$ (Latent Overlap): for some $c>0$,
\[ c<\pr(W_{it}=1|\alpha_i,\bbeta_t)<1-c.\]
\end{assumption}
\begin{remark}
If we strengthened the condition to require only conditioning on the unit component,
\[ W_{it}\ \indep\ Y_{it}(0)\ \Bigl|\ \aalpha_i,\]
the problem would be straightforward. In that case we could immediately use the average outcome for the treated unit during control periods as an estimator for the missing potential outcome. If a treated unit period pair is $(i,t)$ the imputed value would be
\[ \hat Y_{it}(0)=\frac{1}{T-1}\sum_{s\neq t}Y_{is}.\]
In that case we would discard observations on units other than the treated unit for the purposes of estimating $Y_{it}(0)$.

Similarly the problem would be simple to solve if we strengthened the condition to
\[ W_{it}\ \indep\ Y_{it}(0)\ \Bigl|\ \bbeta_t.\]

The fundamental challenge we address in this paper is that the Latent Factor Unconfoundedness condition requires conditioning on both $\aalpha_i$ and $\bbeta_t$, and we do not observe  either of them.
Of course this challenge arises exactly from the concern that units  differ in unobserved but relevant aspects, and time periods differ in unobserved but relevant aspects, and which is often the motivation for collecting panel data.
\end{remark}
\begin{remark}
Latent Factor Unconfoundedness has no testable implication. This can be seen directly by considering the case where we observe $\alpha_i$ and $\beta_t$. In that case Latent Factor Unconfoundedness is directly equivalent to a standard unconfoundedness assumption when we view the unit of analysis the unit/time-period pair. 
\end{remark}
\begin{remark}
We do not restrict the dependence of the assignment on the unit and time components.
\end{remark}

Here we want to extend the insights from the previous section to the case where we are interested in estimating the average treatment effect for the treated, $\tau=\sum_{i,t} W_{it}(Y_{it}(1)-Y_){it}(0))/\sum_{i,t} W_{it}.$ This involves estimating $\mu_{it}=\mathbb{E}[Y_{it}(0)|\aalpha_i,\bbeta_t]$ for all treated unit/period pairs. The primary challenge  relative to estimating $\mu_{it}$ in the previous section is that we cannot estimate the marginal covariance between outcomes for units $i$ and $k$ because we do not observe $Y_{it}(0)$ and $Y_{kt}(0)$ for all units and periods. Moreover, the time units and time periods where we do observe $Y_{it}(0)$ may be systematically different from those where we do not do so. To deal with that we modify the definition of the sets ${\cal J}^*(\aalpha)$ and ${\cal J}_{m,\nu}(i)$. To avoid notational confusion we refer to the new sets
as ${\cal C}^*(\aalpha)$ and ${\cal C}_{m,\nu}(i)$ (where the ${\cal C}$ stands for Causal).

Define the propensity score and the marginal treatment assignment probability
\[ e(\aalpha_i,\bbeta_t)\equiv\pr(W_{it}=1|\aalpha_i,\bbeta_t). \]
and the set
\[ {\cal C}^*(\aalpha)\equiv 
\left\{
\aalpha'\in[0,1],\st 
\sup_{\alpha''}
\biggl|\mathbb{E}_\beta\left[\left.\left\{\mu(\aalpha,\bbeta)-\mu(\aalpha',\bbeta)\right\}\mu(\aalpha'',\bbeta)\right|\aalpha,\aalpha',\aalpha''\right]
\biggr|=0
\right\}
.\]

First define for a subset  ${\cal N}$ of $\{1,\ldots,N_m\}$ the set of periods 
\[{\cal S}({\cal N})=\{t=1,\ldots,T_m|
W_{jt}=0\forall j\in{\cal N}\}\] where all the units in the set ${\cal N}$ are in the control group. By the overlap assumption the cardinality of the set ${\cal S}({\cal N})$ will be increasing as $T_m$ increases for a finite set ${\cal N}.$
For a given unit $i$ we now want to construct sets ${\cal J}_{m,\nu}(i)$ such that if $j\in{\cal J}_{m,\nu}(i) $, then $\mu(\aalpha_i,\beta)$ is close to $\mu(\aalpha_j,\beta)$  for all $\bbeta$. 
Previously we operationalized this by ensuring that the expectation of $(\mu(\aalpha_i,\bbeta_t)-\mu(\aalpha_j,\bbeta_t)^2$ is close to zero. This in turn we ensured by requiring that the  expectation of $\mu(\aalpha_i,\beta_t)-\mu(\aalpha_j,\beta_t))\mu(\aalpha_k,\bbeta_t)$  is close to zero for all other units $k$.  
In fact we only used the fact that this expetation was close to zero at two choices for $k$: at $k=i$ and at $k=j$. Now to deal with the missing $Y_{it}(0)$ we modify this by ensuring that the {\it conditional} expectation of $\mu(\aalpha_i,\beta_t)-\mu(\aalpha_j,\beta_t))\mu(\aalpha_k,\bbeta_t)$  is close to zero for all other units $k$. 
The conditioning is designed to ensure that we can compare the expectations for two choices of $k$. This means that we now require that for all pairs $(k,l)$ the expectations 
$(\mu(\aalpha_i,\beta_t)-\mu(\aalpha_j,\beta_t))\mu(\aalpha_k,\bbeta_t)$  
and $(\mu(\aalpha_i,\beta_t)-\mu(\aalpha_j,\beta_t))\mu(\aalpha_l,\bbeta_t)$  are close to zero, {\it conditional} on $W_{it}=W_{jt}=W_{kt}=W_{lt}=0$.

Define
\[ {\cal J}_{m,\nu}(i)\equiv \left\{j=1,\ldots,N_m,\st
\max_{k,l\neq i,j}\left\{
\left|\frac{\sum_{t\in{\cal S}(\{i,j,k,l\})}(Y_{it}-Y_{jt}) Y_{kt}}{
\sum_{t\in{\cal S}(\{i,j,k,l\}} 1}
\right|\right.
\right.\]
\[\left.\left.+ 
\left|\frac{\sum_{t\in{\cal S}(\{i,j,k,l\})}(Y_{it}-Y_{jt}) Y_{lt}}{
\sum_{t\in{\cal S}(\{i,j,k,l\}} 1}
\right|\right\}
\leq \nu
\right\},\]

Then, under suitable rate conditions on $\nu$ we can estimate $\tau$ as
\[ \hat\tau=\frac{\sum_{i,t} W_{it}\Bigl( Y_{it}-\hat Y_{it,\nu}(0)\Bigr)} {\sum_{i,t} W_{it}},\qquad
{\rm where}\quad
 \hat Y_{it,\nu}(0)=\frac{\sum_{j\in{\cal J}_{m,\nu}(i)} Y_{jt}}{\sum_{j\in{\cal J}_{m,\nu}(i)} 1}.\]

\section{Nonparametric Decompositions}
\label{sec:decomp}

In this section, we will change the interpretation of $W_{it}$ to denote group membership, and we do not impose unconfoundedness assumptions. Rather, we make use of identities to break out differences in group means into several components.  We further modify notation slightly and assume that there is a parallel NPM for units where $W_{it}=1$, so that for $w\in\{0,1\}$: 
\begin{equation}\label{eq:modelgen} Y_{it}(w)=g_w(\alpha_i^w,\beta_t^w,\varepsilon^w_{it})\end{equation}
where the unobserved components  $\aalpha_i^w$, $\beta_t^w$ and $\varepsilon^w_{it}$ may be vector valued. We further let

\begin{equation}\label{eq:modelmu}
\mu^w(\aalpha_i^w,\bbeta_t^w)=\mathbb{E}[Y_{it}|\aalpha_i^w,\bbeta_t^w].
\end{equation}

Then, we develop a decomposition by first using iterated expectations and substituting in (\ref{eq:modelmu}), and then adding and subtracting a term:

\begin{center}
\begin{align}
\label{eq:diff}
\mathbb{E}\bigl[Y_{it}|W_{it}=1\bigr]-\mathbb{E}\bigl[Y_{it}|W_{it}=0\bigr] 
&=\mathbb{E}_{\aalpha_i^1,\bbeta_t^1}\bigl[\mu^1(\aalpha_i^1,\bbeta_t^1)\bigr]-
\mathbb{E}_{\aalpha_i^0,\bbeta_t^0}\bigl[\mu^0(\aalpha_i^0,\bbeta_t^0)\bigr]\\
\label{eq:decompunexpmu}
&=\mathbb{E}_{\aalpha_i^1,\bbeta_t^1}\bigl[\mu^1(\aalpha_i^1,\bbeta_t^1)\bigr]- \mathbb{E}_{\aalpha_i^1,\bbeta_t^1}\bigl[\mu^0(\aalpha_i^1,\bbeta_t^1)\bigr]\\ \label{eq:decompexpmu}
&+\mathbb{E}_{\aalpha_i^1,\bbeta_t^1}\bigl[\mu^0(\aalpha_i^1,\bbeta_t^1)\bigr]-
\mathbb{E}_{\aalpha_i^0,\bbeta_t^0}\bigl[\mu^0(\aalpha_i^0,\bbeta_t^0)\bigr]
\end{align}
\end{center}

If $\aalpha_i$ and $\bbeta_t$ were observed, the analysis above would be an example of standard decompositions (e.g. \cite{blau2017gender}), where(\ref{eq:decompunexpmu}) and (\ref{eq:decompexpmu}) are referred to as the ``unexplained'' and ``explained'' gap, respectively. The unexplained gap is the part that remains when the two groups have the same distribution of observables, while the explained gap is the difference that arises when both groups have the same expected outcome function (but different distributions of characteristics). In our case, $\aalpha_i$ and $\bbeta_t$ are unobserved, but the analysis of Section \ref{sec:ATE} can be adapted to show that (\ref{eq:decompunexpmu}) and (\ref{eq:decompexpmu}) are identified and can be estimated under the overlap condition.

\section{Conclusion}

We propose a nonparametric version of a factor model for estimating average treatment effects in a panel data setting. The model substantially generalizes the commonly used linear factor models and allows for a clear statement of the critical assumptions. 

We establish that there is a consistent estimator of the expected non-treated outcome for any given unit and time period, and we use this result to provide conditions under which the average treatment effect is identified. 

Our results can also be applied in other settings, such as to problems of decomposing gaps in outcomes by group, as is common in the literature on the gender wage gap.

\newpage

\appendix

\setcounter{equation}{0}
\renewcommand{\theequation}{A\arabic{equation}}

\setcounter{lemma}{0}
\renewcommand{\thelemma}{A\arabic{lemma}}

\section{Proofs and Additional Results}

\noindent{\sc Proof of Lemma \ref{lemma_main}:}
By the definition of ${\cal J}^*(\aalpha)$, if $\aalpha'\in{\cal J}^*(\aalpha)$ then
\[\mathbb{E}_\beta\left[\left.\left\{\mu(\aalpha,\beta)-\mu(\aalpha',\beta)\right\}\mu(\alpha'',\beta)\right|\aalpha,\aalpha',\aalpha''\right]=0
\]
for all $\alpha''$, so
substituting $\alpha''=\alpha$ implies
\[\mathbb{E}_\beta\left[\left.\left\{\mu(\aalpha,\beta)-\mu(\aalpha',\beta)\right\}\mu(\alpha,\beta)\right|\aalpha,\aalpha'\right]=0
\]
\[\Longrightarrow\quad \mathbb{E}_\beta[\mu(\aalpha,\beta)^2|\aalpha]
=\mathbb{E}_\beta[\mu(\aalpha,\beta)\mu(\aalpha',\bbeta)|\aalpha,\aalpha'].
\]
Also, by substituting $\alpha''=\alpha'$ we have
\[\mathbb{E}_\beta\left[\left.\left\{\mu(\aalpha,\beta)-\mu(\aalpha',\beta)\right\}\mu(\alpha',\beta)\right|\aalpha,\aalpha'\right]=0
\]
\[\Longrightarrow\quad \mathbb{E}_\beta[\mu(\aalpha,\beta)\mu(\aalpha',\bbeta)|\aalpha,\aalpha']=\mathbb{E}_\beta[\mu(\aalpha',\beta)^2|\aalpha'].
\]
Thus
\[\mathbb{E}_\beta[\mu(\aalpha,\beta)^2|\aalpha,\aalpha']=\mathbb{E}_\beta[\mu(\aalpha,\beta)\mu(\aalpha',\bbeta)|\aalpha,\aalpha]
=\mathbb{E}_\beta[\mu(\aalpha',\beta)^2|\alpha,\aalpha']
\]
\[\Longrightarrow\quad \mathbb{E}[(\mu(\alpha,\beta)-\mu(\alpha',\beta))^2|\alpha,\aalpha']\]
\[=\mathbb{E}_\beta[\mu(\aalpha,\beta)^2|\aalpha,\aalpha']-2\mathbb{E}_\beta[\mu(\aalpha,\beta)\mu(\aalpha',\bbeta)|\aalpha,\aalpha']
+\mathbb{E}_\beta[\mu(\aalpha',\beta)^2|\alpha,\aalpha']=0.\]
$\square$

\begin{lemma}
    Suppose Assumptions \ref{assumption:model} and \ref{assumption:smoothness} hold. Then $\mu(\aalpha,\bbeta)$ is is differentiable with bounded first derivatives.
\end{lemma}

Define
\[ {\cal J}^1_{m,\nu}(i)\equiv \left\{j=1,\ldots,N_m,\st
\max_{k}
\left| \mathbb{E}_{\bbeta_t}\left[\left.\left\{\mu(\aalpha_i,\bbeta_t) -\mu(\aalpha_j,\bbeta_t)\right\}\mu(\aalpha_k,\bbeta_t)\right|\aalpha_i,\aalpha_j,\aalpha_k\right]\right|\leq \nu
\right\}.\]


\begin{lemma}\label{lemma2a}
Suppose Assumptions \ref{assumption:model}, \ref{assumption:asymptotic}, and \ref{assumption:smoothness}
hold. For any $\epsilon,\xi>0$, if
\[N_m\geq\frac{\ln(\epsilon\xi/(8\overline{\mu'}\overline{\mu}))}
{\ln(1-\xi/(8\overline{\mu'}\overline{\mu}))},\] then
\[ \pr_{\aalpha_j}\left(
\left.
\mathbb{E}_{\beta_t}\left[\left.\left\{\mu(\alpha_i,\beta_t)-\mu(\alpha_j,\beta_t)\right\}^2\right|\aalpha_i,\aalpha_j,j\in {\cal J}^1_{m,\xi/4}(i) \right]>\xi
\right|\aalpha_i,j\in {\cal J}^1_{m,\xi/4}(i)\right)<\epsilon.\]
Here the expectation conditional on $j\in{\cal J}^1_{m,\xi/4}(i)  $ is to be understood to be zero if the conditioning set is empty.
\end{lemma}
\noindent{\sc Proof of Lemma \ref{lemma2a}:
}
Define $\eta=\xi/(8\overline{\mu'}\overline{\mu})$ and $\nu=\xi/4.$
Partition the set $[0,1]$ into $R$ subsets  of the  type $((r-1)/R,r/R]$ with $R>1/\eta=(8\overline{\mu'}\overline{\mu})/\xi$.
Let $A_{1m}$ be the event that there is for each $r$ a unit $j$ such that $\aalpha_j\in((r-1)/R,R]$, and let $A_{1m}^c$ be its complement.
The probability of the event  $A_{1m}^c$ is less than $R(1-\eta)^{N_m}$, which is less than $\epsilon$ by the assumption on $N_m.$

Now suppose $j\in {\cal J}^1_{m,\xi/4}(i).$ Then for any $\aalpha\in[0,1]$, conditional on the event $A_{1m}$,
there is a $k$ such that $|\aalpha-\aalpha_k|<\eta.$
Because $j\in{\cal J}^1_{m,\xi/4}(i)$, it follows that
\[\biggl| \mathbb{E}_{\bbeta_t}\left[\left.\left\{\mu(\aalpha_i,\bbeta_t) -\mu(\aalpha_j,\bbeta_t)\right\}\mu(\aalpha_k,\bbeta_t)\right|\aalpha_i,\aalpha_j,\aalpha_k\right]\biggr|\leq \xi/4.\]
Hence
\[\left| \mathbb{E}_{\bbeta_t}\left[\left.\left\{\mu(\aalpha_i,\bbeta_t) -\mu(\aalpha_j,\bbeta_t)\right\}\mu(\aalpha,\bbeta_t)\right|\aalpha_i,\aalpha_j,\aalpha\right]\right|\]
\[\biggl| \mathbb{E}_{\bbeta_t}\left[\left\{\mu(\aalpha_i,\bbeta_t) -\mu(\aalpha_j,\bbeta_t)\right\}\mu(\aalpha_k,\bbeta_t)
+\biggl\{\mu(\aalpha_i,\bbeta_t) -\mu(\aalpha_j,\bbeta_t)\right\}\left\{\mu(\aalpha,\bbeta_t)-\mu(\aalpha_k,\bbeta_t)\biggr\}
\biggl|\aalpha_i,\aalpha_j,\aalpha_k,\aalpha \right]\biggr|\]
\[\leq 
\left| \mathbb{E}_{\bbeta_t}\left[\left.\left\{\mu(\aalpha_i,\bbeta_t) -\mu(\aalpha_j,\bbeta_t)\right\}\mu(\aalpha_k,\bbeta_t)
\right|\aalpha_i,\aalpha_j,\aalpha_k,\aalpha\right]
\right|\]
\[
+\biggl| \mathbb{E}_{\bbeta_t}\left[\left.
\left.\left\{\mu(\aalpha_i,\bbeta_t) -\mu(\aalpha_j,\bbeta_t)\right\}\left\{\mu(\aalpha,\bbeta_t)-\mu(\aalpha_k,\bbeta_t)\right\}
\right|\aalpha_i,\aalpha_j,\aalpha_k,\aalpha\right]
\right|
\]
\[\leq \xi/4+2\overline{\mu}\overline{\mu'}\eta= \xi/2.\]
Substituting $\aalpha=\aalpha_i$ implies
\[\left| \mathbb{E}_{\bbeta_t}\left[\left.\mu^2(\aalpha_i,\bbeta_t)\right|\aalpha_i\right]
-\mathbb{E}_{\bbeta_t}\left[\left.\mu(\aalpha_j,\bbeta_t)\mu(\aalpha_i,\bbeta_t)\right|\aalpha_i,\aalpha_j\right]
\right|<\xi/2.\]
Substituting $\aalpha=\aalpha_j$ implies
\[\left| \mathbb{E}_{\bbeta_t}\left[\left.\mu^2(\aalpha_j,\bbeta_t)\right|\aalpha_i\right]
-\mathbb{E}_{\bbeta_t}\left[\left.\mu(\aalpha_j,\bbeta_t)\mu(\aalpha_i,\bbeta_t)\right|\aalpha_i,\aalpha_j\right]
\right|<\xi/2,\]
which in combination implies
\[
\mathbb{E}_\beta\left[\left.\left\{\mu(\alpha_i,\beta)-\mu(\alpha_j,\beta)\right\}^2\right|\aalpha_i,\aalpha_j,j\in {\cal J}^1_{m,\xi/4}(i) ,A_{1m}\right]<\xi .
\]
Because the probability of the event $A_{1m}$ is at least $1-\epsilon,$ the proof is complete.
$\square$

Next, define
\[ {\cal J}^2_{m,\nu}(i)\equiv \left\{j=1,\ldots,N_m,\st
\max_{k\neq i,j}
\left| \frac{1}{T_m}\sum_{t=1}^{T_m}\left\{\mu(\aalpha_i,\bbeta_t) -\mu(\aalpha_j,\bbeta_t)\right\}\mu(\aalpha_k,\bbeta)\right|\leq \nu
\right\},\]


\begin{lemma}\label{lemma2b}
Suppose Assumptions \ref{assumption:model}, \ref{assumption:asymptotic}, and \ref{assumption:smoothness} 
hold. For any $\epsilon,\xi>0$, if
\[N_m\geq\frac{\ln(\epsilon\xi/(16\overline{\mu'}\overline{\mu})}{\ln(1-\xi/16\overline{\mu'}\overline{\mu})},\qquad{\rm and}\quad T_m\geq \frac{256N_m^2 C^2_Y}{\epsilon^2\xi^2},\]
then
\[ \pr\left(\mathbb{E}_\beta\left[\left.\left\{\mu(\alpha_i,\beta)-\mu(\alpha_j,\beta)\right\}^2\right|\aalpha_i,\aalpha_j,j\in {\cal J}^2_{m,\xi/8}(i) \right]>\xi \right)<\epsilon.\]
Here the expectation conditional on $j\in{\cal J}^1_{m,\xi/8}(i)  $ is to be understood to be zero if the conditioning set is empty.
\end{lemma}
\noindent{\sc Proof of Lemma \ref{lemma2b}:
}
Now there are two events we wish to condition on. The first, denoted by $A_{1m}$, is again the event that there are units $k$ with $\aalpha_k$ in each of the sets $((r-1)/R,r/R]$ for all $r=1,\ldots,R,$ for $R\geq (16\overline{\mu'}\overline{\mu})/\xi$ as in the proof of Lemma \ref{lemma2a}. 
This event has probability more than $1-\epsilon/2$ by the condition on $N_m.$
The second event, denoted by $A_{2m},$ is the event that 
\[\max_{i,j}
\left|
\frac{1}{T_m}
\sum_{t=1}^{T_m} 
\left\{
\mu(\alpha_j,\beta_t)\mu(\alpha_k,\beta_t)-
\mathbb{E}_\beta
\left[\left.\mu(\alpha_j,\beta_t)\mu(\alpha_k,\beta_t)\right| \aalpha_i,\aalpha_j\right]
\right\}
\right|<\xi/16.\]
The event $A_{2m}$ has probability more than $1-\epsilon/2$ by Chebyshev's inequality given the restriction on $T_m$ (where $C_Y$ is an upper limit on $|g(\aalpha,\bbeta,\varepsilon)|$).
Conditional on the event $A_{2m}$ $j$ being in the set ${\cal J}^2_{m,\xi/8}(i)  $  implies
\[ \max_{k}
\biggl| \mathbb{E}\biggl[\biggl\{\mu(\aalpha_i,\bbeta_t) -\mu(\aalpha_j,\bbeta_t)\biggr\}\mu(\aalpha_k,\bbeta)\biggr|\]
\[ \leq
\max_{k}
\left| \frac{1}{T_m}\sum_{t=1}^{T_m}\left\{\mu(\aalpha_i,\bbeta_t) -\mu(\aalpha_j,\bbeta_t)\right\}\mu(\aalpha_k,\bbeta)\right|\]
\[+
\max_{k}
\biggl| \frac{1}{T_m}\sum_{t=1}^{T_m}\left\{\mu(\aalpha_i,\bbeta_t) -\mu(\aalpha_j,\bbeta_t)\right\}\mu(\aalpha_k,\bbeta)-
\mathbb{E}\biggl[\left\{\mu(\aalpha_i,\bbeta_t) -\mu(\aalpha_j,\bbeta_t)\right\}\mu(\aalpha_k,\bbeta)\biggr|\]
\[\leq \xi/4\]
and thus 
$j\in{\cal J}^1_{m,\xi/4}(i)  $. The remainder follows from an application of Lemma \ref{lemma2a}.
$\square$

\begin{lemma}
    \label{lemma2c}
Suppose Assumptions \ref{assumption:model}, \ref{assumption:asymptotic}, and \ref{assumption:smoothness}
hold.
Then 
for 
\[T_m\geq\frac{N_m^2 C^2_Y}{\epsilon\nu^2}\]
\[ \pr\left(
\max_{i\neq j} 
\left|
\frac{1}{T_m}
\sum_{t=1}^{T_m} Y_{it}Y_{jt}-
\mathbb{E}
\left[\left.\mu(\aalpha_i,\beta_t)\mu(\alpha_j,\bbeta_t)\right| \aalpha_i,\aalpha_j\right]
\right|>\nu
\right)<\epsilon\]
\end{lemma}
\noindent{\sc Proof of Lemma \ref{lemma2c}:}
For a fixed $i$ and $j$, we have
\[ \pr\left(
\left|
\frac{1}{T_m}
\sum_{t=1}^{T_m} Y_{it}Y_{jt}-
\mathbb{E}
\left[\left.\mu(\aalpha_i,\beta_t)\mu(\alpha_j,\bbeta_t)\right| \aalpha_i,\aalpha_j\right]
\right|>\nu
\right)<\frac{C^2_Y}{T_m \xi^2}.\]
Because there are less than $N_m^2$ pairs $(i,j)$, it follows that
\[ \pr\left(
\max_{i\neq j} 
\left|
\frac{1}{T_m}
\sum_{t=1}^{T_m} Y_{it}Y_{jt}-
\mathbb{E}
\left[\left.\mu(\aalpha_i,\beta_t)\mu(\alpha_j,\bbeta_t)\right| \aalpha_i,\aalpha_j\right]
\right|>\nu
\right)<\frac{N_m^2C^2_Y}{T_m \nu^2}.\]
For $T_m$ satisfying the condition in the lemma the right hand side is less than $\epsilon.$
$\square$

\noindent{\sc Proof of Theorem \ref{theorem1}:}
In addition to the events $A_{1m}$ and $A_{2m}$ defined in the proofs of Lemmas \ref{lemma2a}
 and \ref{lemma2b},
 define the event $A_{3m}$ as the event that
\[ 
\max_{i\neq j} 
\left|
\frac{1}{T_m}
\sum_{t=1}^{T_m} Y_{it}Y_{jt}-
\mathbb{E}
\left[\left.\mu(\aalpha_i,\beta_t)\mu(\alpha_j,\bbeta_t)\right| \aalpha_i,\aalpha_j\right]
\right|<\nu.\]
For fixed $i$ and $j$, with $i\neq j$
the probability of the event 
\[\left|
\frac{1}{T_m}
\sum_{t=1}^{T_m} Y_{it}Y_{jt}-
\mathbb{E}
\left[\left.\mu(\aalpha_i,\beta_t)\mu(\alpha_j,\bbeta_t)\right| \aalpha_i,\aalpha_j\right]
\right|<\nu\]
is less than $C^2_Y/(T_m\nu^2).$ Hence the probability of the complement of $A_{3m}$,
\[ 
\max_{i\neq j} 
\left|
\frac{1}{T_m}
\sum_{t=1}^{T_m} Y_{it}Y_{jt}-
\mathbb{E}
\left[\left.\mu(\aalpha_i,\beta_t)\mu(\alpha_j,\bbeta_t)\right| \aalpha_i,\aalpha_j\right]
\right|>\nu\]
is less than $N_m^2 C^2_Y/(T_m\nu^2).$

The probability of events $A_{1m}$, $A_{2m}$, and $A_{3m}$ jointly holding is at least $1-\varepsilon.$ Conditional on all three events,
\[ \mathbb{E}_\beta\left[\left.\left\{\mu(\alpha_i,\beta)-\mu(\alpha_j,\beta)\right\}^2\right|\aalpha_i,\aalpha_j,j\in {\cal J}_{m,\nu_m}(i), A_{1m}, A_{2m}, A_{3m} \right]<\xi,\]
which proves the result.
$\square$

\newpage
\bibliography{\bib}

\vfill\eject
\end{document}